\documentclass[floatfix,aps,superscriptaddress,prd,amsmath,nofootinbib,preprintnumbers,onecolumn]{revtex4}

\usepackage{graphicx}
\usepackage{bm}
\usepackage{float}
\usepackage[
colorlinks=true,        
citecolor=blue,         
linkcolor=blue,         
urlcolor=blue           
]{hyperref}             

\newcommand{\nc}{\newcommand*}
\nc{\Om}{\Omega}
\nc{\ogw}{\Omega_{\mathrm{GW}}}
\nc{\rd}{\mathrm{d}}
\nc{\eg}{\textit{e.g.~}}
\nc{\red}[1]{\textcolor{red}{#1}}
\nc{\lvc}{LIGO/Virgo} 

\def\({\left(}
\def\){\right)}
\def\[{\left[}
\def\]{\right]}

\def\e{\begin{equation}}
\def\q{\end{equation}}
\def\m{\begin{eqnarray}}
\def\n{\end{eqnarray}}

\begin{document}

\title{Measuring the primordial curvature perturbations from the scalar induced gravitational waves}

\author{Jun Li}
\email{lijun@qust.edu.cn}
\affiliation{School of Mathematics and Physics,
    Qingdao University of Science and Technology,\\
    Qingdao 266061, China}
\affiliation{CAS Key Laboratory of Theoretical Physics,\\
    Institute of Theoretical Physics, Chinese Academy of Sciences,\\
    Beijing 100190, China}

\author{Guang-Hai Guo}
\email{thphys$_$qust@163.com}
\affiliation{School of Mathematics and Physics,
    Qingdao University of Science and Technology,\\
    Qingdao 266061, China}

\date{\today}

\begin{abstract}
The scalar induced gravitational waves are produced from primordial curvature perturbations in the second order of perturbations. We constrain the fractional energy density of scalar induced gravitational waves from gravitational waves observations. If there is no detection of the scalar induced gravitational waves, the fractional energy density of scalar induced gravitational waves is constrained by some upper limits. Depends on these upper limits, we can obtain the constraints on the power spectrum of the primordial curvature perturbations. For a power-law scalar power spectrum, the constraints on the power spectrum are affected by adding the upper limit of scalar induced gravitational waves from Square Kilometer Array (SKA). In the standard model, the mean values of the scalar amplitude and the spectral index shift to lower values when SKA is added to the combination of Cosmic Microwave Background (CMB) and Baryon Acoustic Oscillation (BAO) datasets, namely $\ln\(10^{10}A_s\)=3.038\pm0.013$ and $n_s=0.9589^{+0.0021}_{-0.0011}$ at $68\%$ confidence level. We also consider the effects of the existing
ground-based gravitational-wave detectors, the existing Pulsar Timing Arrays (PTAs) and Five-hundred-meter Aperture Spherical radio Telescope (FAST), while the constraints from CMB+BAO datasets are totally within their upper limits of scalar induced gravitational waves. Furthermore, we characterize the scalar fluctuation spectrum in terms of the spectral index $n_s$ and its first two derivatives. We calculate corresponding power spectrum of scalar induced gravitational waves theoretically and give the constraints on the running of the spectral index and the running of the running of the spectral index.

\end{abstract}

\maketitle

\section{introduction}
The cosmological perturbation theory has been developed so fast these decades owning to cosmological observations, such as the Cosmic Microwave Background (CMB). The primordial perturbations affect not only temperature but also polarization by the  scalar and tensor perturbations \cite{Riotto:2002yw, Cabella:2004mk, Zaldarriaga:1996xe, Ma:1995ey}.  The polarization can be decomposed into E-mode and B-mode, while the
B-mode component mainly comes from the tensor perturbation on very large scales and encodes the information of primordial gravitational waves. The upper limit on the tensor-to-scalar ratio is $r_{0.05}<0.038$ at
$95\%$ confidence level from the combinations of Planck satellite \cite{Planck:2018vyg}, the BICEP/Keck Observations \cite{BICEP:2021xfz} and the Baryon Acoustic Oscillation (BAO) \cite{Beutler:2011hx,Ross:2014qpa,BOSS:2016wmc} which reflects the constraint on primordial gravitational waves.

Besides the primordial gravitational waves, the scalar induced gravitational waves are generated in the second order of perturbations. The curvature perturbations couple to the tensor perturbations at second-order which produce the induced gravitational waves in the radiation dominated era. Although the induced gravitational waves are suppressed by the square of curvature perturbations, but they can compare with primordial gravitational waves if the curvature perturbations are large enough. The enhancement of induced gravitational waves can be realized in some models of gravity \cite{Carbone:2004iv,Matarrese:1997ay,Noh:2004bc}, inflation \cite{Alabidi:2012ex, Zhou:2020kkf,Espinosa:2018eve,Osano:2006ew,Alabidi:2013lya,Orlofsky:2016vbd,Di:2017ndc,Cai:2019jah,Tada:2019amh,Inomata:2019zqy,Inomata:2019ivs,Xu:2019bdp} or scalar power spectrum \cite{Assadullahi:2009nf,Ananda:2006af,Kohri:2018awv,Lu:2019sti,Yuan:2019wwo,Cai:2019elf,Chen:2019xse,Baumann:2007zm,Inomata:2016rbd,Assadullahi:2009jc,Giovannini:2010tk,Cai:2018dig,Unal:2018yaa,Bartolo:2018rku,Inomata:2018epa,Cai:2019amo}. The evolution of induced gravitational waves in the radiation-dominated era was studied \cite{Ananda:2006af,Baumann:2007zm}. As examples they computed the gravitational wave background generated by both a power-law spectrum on all scales, and a delta-function power spectrum on a single scale which were confirmed numerically and analytically by \cite{Osano:2006ew,Assadullahi:2009jc,Giovannini:2010tk,Alabidi:2012ex,Alabidi:2013lya,Inomata:2016rbd,Orlofsky:2016vbd,Di:2017ndc,Kohri:2018awv}. The curvature perturbations at small scales were constrained from the induced gravitational waves by gravitational-wave projects \cite{Inomata:2018epa}. Gravitational waves induced by non-Gaussian scalar perturbations were evaluated \cite{Cai:2018dig}, which also forecast a distinctive observational perspective. The curvature perturbations originated from the hypothetical existence of primordial black holes were considered \cite{Cai:2019elf,Chen:2019xse}. The corresponding induced gravitational waves was detectable by current and future observations.

The detection of scalar induced gravitational waves becomes important in cosmological perturbation theory. The gravitational waves detections provide the latest way to find scalar induced gravitational waves which include Laser Interferometer Gravitational-wave Observatory (LIGO) and Virgo detector \cite{LIGOScientific:2019vic}, Laser Interferometer Space Antenna (LISA) detector \cite{Caprini:2015zlo}, International Pulsar Timing Array (IPTA) \cite{Verbiest:2016vem}, Five-hundred-meter Aperture Spherical radio Telescope (FAST) \cite{Nan:2011um, Kuroda:2015owv} and Square Kilometer Array (SKA) \cite{Kuroda:2015owv}. IPTA is the combination of three Pulsar Timing Array (PTA) projects \cite{Hellings:1983fr}, namely European Pulsar Timing Array (EPTA) \cite{Desvignes:2016yex}, Parkes Pulsar Timing Array (PPTA) \cite{Hobbs:2013aka} and North American Observatory for Gravitational Waves (NANOGrav) \cite{McLaughlin:2013ira}. All of these detectors are sensitive to the fractional energy density which may contain information of scalar induced gravitational waves. If there is no detection of scalar induced gravitational waves, the fractional energy density of scalar induced gravitational waves is constrained by some upper limits. For LIGO and Virgo detector, the upper limit is $\Omega_{\mathrm{GW}}<10^{-7}$ at frequency around $40$ Hz, which is actual observational constraints. For LISA detector, the upper limit is $\Omega_{\mathrm{GW}}<10^{-12}$ at frequency around $10^{-3}$ Hz. For IPTA detector, the upper limit is $\Omega_{\mathrm{GW}}<10^{-16}$ at frequency $1.58*10^{-9}$ Hz. For FAST detector, the upper limit is $\Omega_{\mathrm{GW}}<10^{-19}$ at frequency $6.34*10^{-10}$ Hz. For SKA detector, the upper limit is $\Omega_{\mathrm{GW}}<10^{-22}$ at frequency $3.17*10^{-10}$ Hz. Besides LIGO and Virgo, other limits are forecast. Depends on these upper limits, we can obtain the constraints on the power spectrum of the primordial curvature perturbations. In this paper, we consider a power-law scalar power spectrum and constrain the power spectrum from the upper limits of scalar induced gravitational waves.

\section{the scalar induced gravitational waves}
In the conformal Newtonian gauge, the metric about the Friedmann-Robert-Walker background is taken as
\e
\mathrm{d}s^2=a^2\left\{-(1+2\Phi)\mathrm{d}\eta^2+\left[(1-2\Phi)\delta_{ij}+\frac{h_{ij}}{2}\right]\mathrm{d}x^i\mathrm{d}x^j \right\},      \label{metric}
\q
where $\eta$ is the conformal time, $a(\eta)$ is the scale factor, $\Phi$ is the scalar perturbation and $h_{ij}$ is the tensor perturbation. We neglect the vector perturbation, the first-order gravitational waves and the
anisotropic stress.
In the Fourier space, the tensor perturbation $h_{ij}$ is
\e
h_{ij}(\eta,\mathbf{x})=\int\frac{\mathrm{d}^3k}{(2\pi)^{3/2}}\Bigg(e_{ij}^{+}(\mathbf{k})h_{\mathbf{k}}^+(\eta)+e_{ij}^{\times}(\mathbf{k})h_{\mathbf{k}}^{\times}(\eta)\Bigg)e^{i\mathbf{k}\cdot\mathbf{x}},
\q
where the plus and cross polarization tensors are
\e
e_{ij}^{+}(\mathbf{k})=\frac{1}{\sqrt{2}}\Bigg(e_i(\mathbf{k})e_j(\mathbf{k})-\bar{e}_i(\mathbf{k})\bar{e}_j(\mathbf{k})\Bigg),\quad
e_{ij}^{\times}(\mathbf{k})=\frac{1}{\sqrt{2}}\Bigg(e_i(\mathbf{k})\bar{e}_j(\mathbf{k})+\bar{e}_i(\mathbf{k})e_j(\mathbf{k})\Bigg),
\q
the normalized vectors $e_i(\mathbf{k})$ and $\bar{e}_i(\mathbf{k})$ are orthogonal to each other and to $\mathbf{k}$.
The tensor equation of motion for $h_{ij}$ can be derived straightforwardly from the perturbed Einstein equation up to the second-order. The scalar perturbation couples from tensor perturbation in the second-order equation. The equation for induced gravitational waves with $\Phi_{\mathbf{k}}$ being the source is given by
\e
h_{\mathbf{k}}^{\prime\prime}(\eta)+2\mathcal{H}h_{\mathbf{k}}^\prime(\eta)+k^2h_{\mathbf{k}}(\eta)=4S_{\mathbf{k}}(\eta), \label{gw}
\q
where the prime denotes derivative with respect to conformal time and $\mathcal{H}=a^\prime/a=aH$ is the conformal Hubble parameter. The source term is given by
\e
S_{\mathbf{k}}=\int\frac{\mathrm{d}^3q}{(2\pi)^{3/2}}e_{ij}(\mathbf{k})q_iq_j\Bigg(2\Phi_{\mathbf{q}}\Phi_{\mathbf{k-q}}+\frac{4}{3(1+\omega)}\left(\mathcal{H}^{-1}\Phi^{\prime}_{\mathbf{q}}+\Phi_{\mathbf{q}}\right)\left(\mathcal{H}^{-1}\Phi^{\prime}_{\mathbf{k-q}}+\Phi_{\mathbf{k-q}}\right)\Bigg).
\q
The power spectrum of scalar induced gravitational waves is defined as
\e
\langle h_{\mathbf{k}}(\eta) h_{\mathbf{k}^{\prime}}(\eta)\rangle=\frac{2\pi^2}{k^3}\delta^{(3)}(\mathbf{k}+\mathbf{k}^{\prime})\mathcal{P}_h(\eta, k),
\q
and the fractional energy density is
\e
\Omega_{\mathrm{GW}}(\eta, k)=\frac{1}{24}\Bigg(\frac{k}{aH}\Bigg)^2\overline{\mathcal{P}_h(\eta, k)}.
\q
After calculation, the power spectrum of scalar induced gravitational waves takes the form
\e
\mathcal{P}_h(\eta, k)=4\int_0^\infty\mathrm{d}v\int_{\vert1-v\vert}^{1+v}\mathrm{d}uf^2(v, u, x)\mathcal{P}_{\zeta}(kv)\mathcal{P}_{\zeta}(ku),
\q
where $\mathcal{P}_{\zeta}(k)$ is the power spectrum of the primordial curvature perturbations. The function $f(v, u, x)$ is defined as
\e
f(v, u, x)=I(v, u, x)\frac{4v^2-(1+v^2-u^2)^2}{4vu},
\q
where $I(v, u, x)$ comes from the source term and $x=k\eta$. In the radiation-dominated Universe, the oscillation average is given by \cite{Kohri:2018awv}
\e
\overline{I_{\mathrm{RD}}^2(v, u, x\to\infty)}=\frac{1}{2}\Bigg(\frac{3(u^2+v^2-3)}{4u^3v^3x}\Bigg)^2\Bigg(\Big(-4uv+(u^2+v^2-3)\log\left|\frac{3-(u+v)^2}{3-(u-v)^2}\right|\Big)^2+\pi^2(u^2+v^2-3)^2\Theta(u+v-\sqrt{3})\Bigg),
\q
where $\Theta(u+v-\sqrt{3})$ is the Heaviside theta function.

\begin{figure}[htb]
\centering
\includegraphics[width=12cm]{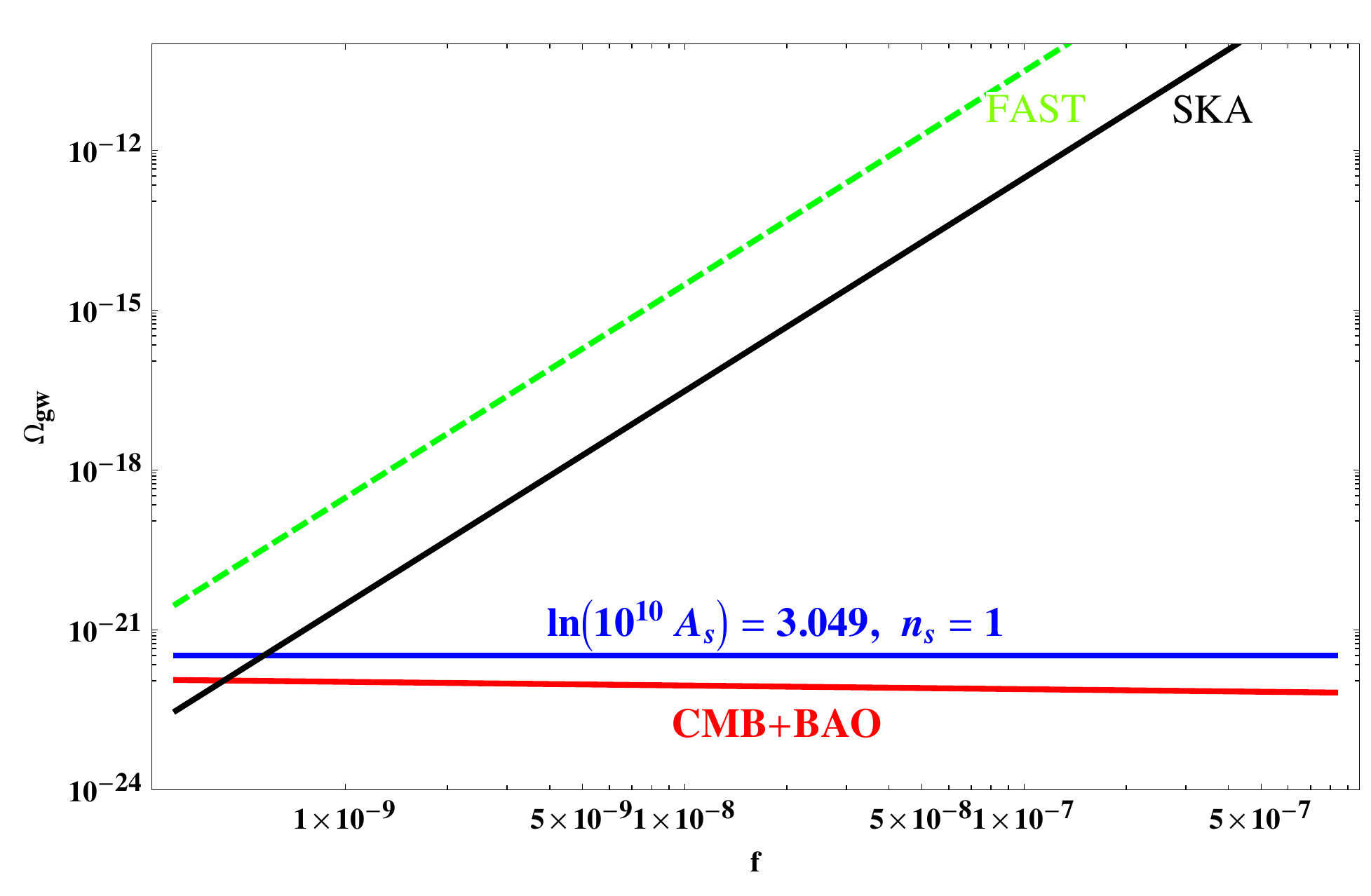}
\caption{The energy density fraction $\Omega_{\mathrm{GW}}$ of the scalar induced gravitational waves from the power-law scalar power spectrum in Eq.~(\ref{igw}) and the sensitivity curves of FAST and SKA detectors. }
\label{figure1}
\end{figure}

For a power-law scalar power spectrum
\e
\mathcal{P}_{\zeta}(k)=A_s\(\frac{k}{k_*}\)^{n_s-1},
\q
the power spectrum of scalar induced gravitational waves is given by \cite{Kohri:2018awv}
\e
\mathcal{P}_h(\eta, k)=\frac{24Q(n_s)}{(k\eta)^2}A^2_s\(\frac{k}{k_*}\)^{2(n_s-1)},
\q
where $A_s$ is the scalar amplitude at the pivot scale $k_*=0.05$ Mpc$^{-1}$, $n_s$ is the scalar spectral index and $Q(n_s)$ is the overall coefficient which depends on $n_s$ and can be found in Table 1 of \cite{Kohri:2018awv}. The fractional energy density becomes
\e
\Omega_{\mathrm{GW}}(\eta, k)=Q(n_s)A^2_s\(\frac{k}{k_*}\)^{2(n_s-1)},  \label{igw}
\q
which corresponds to the quantity evaluated at late times during the radiation dominated era. If it is evaluated today, the present value of the energy fraction is related to the value in the radiation dominated era
\e
\Omega_{\mathrm{GW}}(\eta_0, k)=\Omega_{r,0}\Omega_{\mathrm{GW}}(\eta_c, k), \label{igwtoday}
\q
where $\Omega_{r,0}=\rho_{r,0}/\rho_0$ is the present value of the energy density fraction of radiation and $\eta_c$ is some time after $\Omega_{\mathrm{GW}}(\eta, k)$ has become constant \cite{Inomata:2018cht, Jinno:2013xqa, Saikawa:2018rcs}. According to Planck18\renewcommand{\thefootnote}{\Roman{footnote}}\footnote{Planck18=TTTEEE+lowE+lensing}+BAO observations: $\ln\(10^{10}A_s\)=3.049$ and $n_s=0.9665$, the strength of the scalar induced gravitational waves around the frequency of $10^{-10}$ Hz would be of the order $10^{-22}$ which is presented in Fig.~\ref{figure1}.

The gravitational waves detections also show the sensitivity curves of frequency and $\Omega_{\mathrm{GW}}$ in the detectible ranges which can be used to find scalar induced gravitational waves. If there is no detection of scalar induced gravitational waves, the fractional energy density of scalar induced gravitational waves is constrained by some upper limits. Combine these upper limits with Eq.~(\ref{igw}), we can obtain the constraints on the power spectrum of the primordial curvature perturbations. Compare with LIGO, Virgo, LISA, IPTA and FAST detectors, the sensitivity curve of SKA detector and the energy density fraction $\Omega_{\mathrm{GW}}$ of the scalar induced gravitational waves from the power-law scalar power spectrum in Eq.~(\ref{igw}) would intersect around the frequency of $10^{-10}$ Hz. As the larger amplitude and spectral index of power-law spectrum enhance the fractional energy density, they are more sensitive to the upper limit, such as the $\ln\(10^{10}A_s\)=3.049$ and $n_s=1$ condition. So we could expect that the sensitivity curve of SKA leads to distinct constraints on the amplitude and the scalar spectral index, especially the larger parts. The sensitivity curves of FAST and SKA depend on Table 5 and Figure 4 of Ref.~\cite{Kuroda:2015owv} which compile the sensitivities for FAST and SKA. The observation span is long enough to obtain such wider bandwidth and sensitivity.

\begin{figure}[htb]
\centering
\includegraphics[width=12cm]{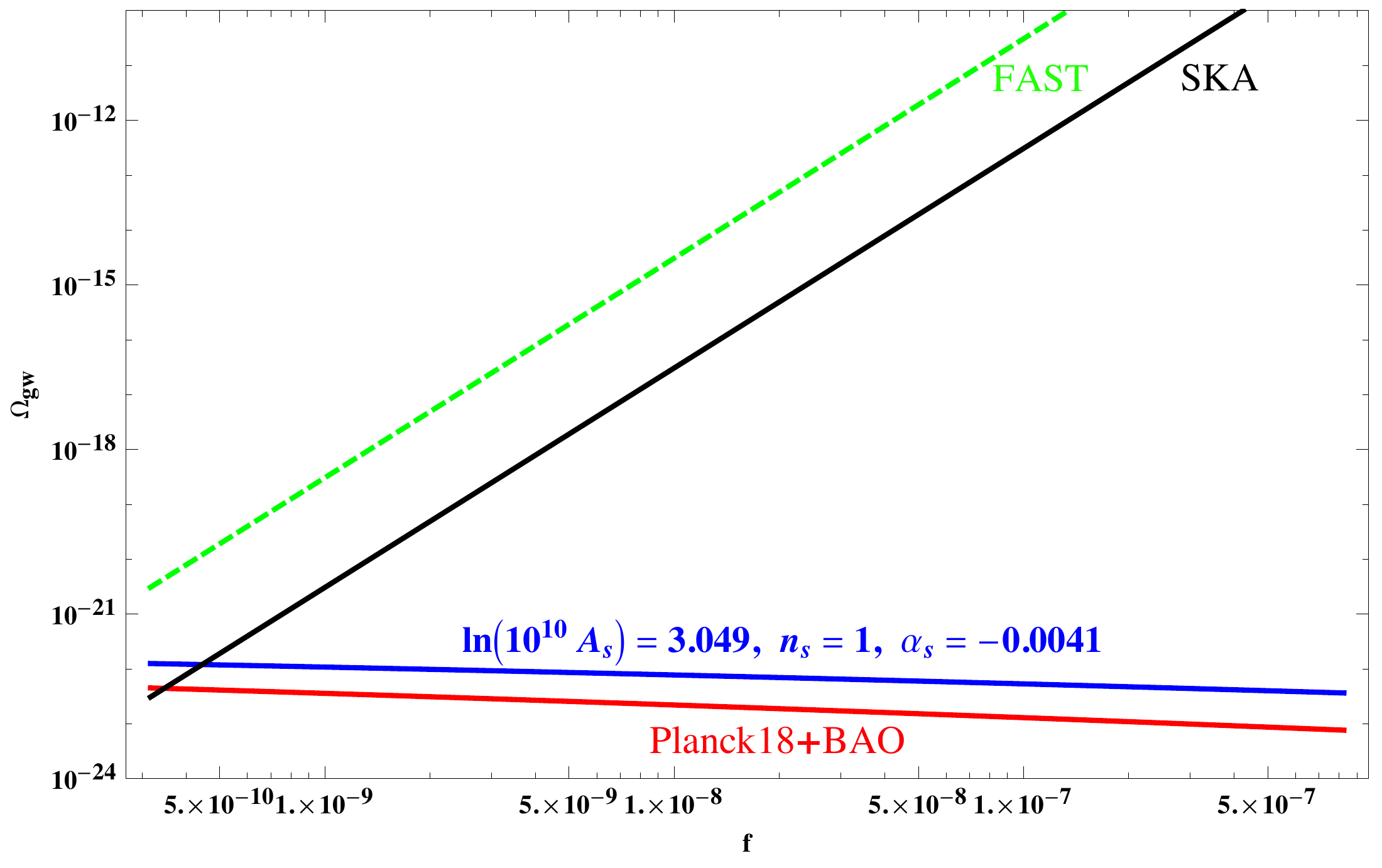}
\caption{The energy density fraction $\Omega_{\mathrm{GW}}$ of the scalar induced gravitational waves in Eq.~(\ref{igw1}) and the sensitivity curves of FAST and SKA detectors. }
\label{figure2}
\end{figure}

Then, we characterize the scalar fluctuation spectrum in terms of the spectral index $n_s$ and its first derivatives with respect to $\ln k$
\m
\mathcal{P}_{\zeta}(k)&=&A_s\(\frac{k}{k_*}\)^{n_s-1+\frac{1}{2}\alpha_s\ln(k/k_*)},
\n
where $\alpha_s\equiv\mathrm{d} n_s/\mathrm{d}\ln k$ is the running of the spectral index. The power spectrum of scalar induced gravitational waves takes the form
\e
\mathcal{P}_h(\eta, k)=\frac{24Q(n_s, \alpha_s, k)}{(k\eta)^2}A^2_s\(\frac{k}{k_*}\)^{2\[{n_s-1+\frac{1}{2}\alpha_s\ln(k/k_*)}\]},
\q
where the overall coefficient is
\m
Q(n_s, \alpha_s, k)=&& \frac{1}{12}\int_0^\infty\mathrm{d}v\int_{\vert1-v\vert}^{1+v}\mathrm{d}u \Bigg(\frac{4v^2-(1+v^2-u^2)^2}{4vu}\Bigg)^2\Bigg(\frac{3(u^2+v^2-3)}{4u^3v^3x}\Bigg)^2\\ \nonumber
&&\Bigg(\Big(-4uv+(u^2+v^2-3)\log\left|\frac{3-(u+v)^2}{3-(u-v)^2}\right|\Big)^2+\pi^2(u^2+v^2-3)^2\Theta(u+v-\sqrt{3})\Bigg)\\ \nonumber
&&\Bigg(\frac{k}{k_*}\Bigg)^{\frac{1}{2}\alpha_s\ln(uv)}\Bigg(uv\Bigg)^{n_s-1+\frac{1}{2}\alpha_s\ln(k/k_*)}v^{\frac{1}{2}\alpha_s\ln{v}}u^{\frac{1}{2}\alpha_s\ln{u}},
\n
which depends on $n_s$, $\alpha_s$ and $k$. The fractional energy density becomes
\e
\Omega_{\mathrm{GW}}(\eta, k)=Q(n_s, \alpha_s, k)A^2_s\(\frac{k}{k_*}\)^{2\[{n_s-1+\frac{1}{2}\alpha_s\ln(k/k_*)}\]}.  \label{igw1}
\q
According to Planck18+BAO observations: $n_s=0.9659$ and $\alpha_s=-0.0041$, the values of the overall coefficient $Q(n_s, \alpha_s, k)$ are shown in Table.~\ref{table1}. The strength of the scalar induced gravitational waves around the frequency of $10^{-10}$ Hz would be of the order $10^{-22}$ which is presented in Fig.~\ref{figure2}. The sensitivity curve of SKA detector and the energy density fraction $\Omega_{\mathrm{GW}}$ of the scalar induced gravitational waves in Eq.~(\ref{igw1}) would intersect around the frequency of $10^{-10}$ Hz. So we could expect that the sensitivity curve of SKA leads to distinct constraints on the running of the spectral index, especially the larger parts.

\begin{table*}[htb]
\newcommand{\tabincell}[2]{\begin{tabular}{@{}#1@{}}#2\end{tabular}}
  \centering
  \begin{tabular}{  c |c| c| c|c|c}
 \hline
 \hline
   mode & \tabincell{c} {$k/k_*=4.1*10^6$} & \tabincell{c}{$k/k_*=8.2*10^6$} & \tabincell{c}{$k/k_*=10$}& \tabincell{c}{$k/k_*=0.5$}& \tabincell{c}{$k/k_*=5.16*10^{17}$}\\
  \hline
  $Q(n_s, \alpha_s, k)$ & $0.8038$ &$0.8035$  &$0.8120$&$0.8143$&$0.7949$\\
  \hline
  \end{tabular}
  \caption{The overall coefficient $Q(n_s, \alpha_s, k)$}
  \label{table1}
\end{table*}

\begin{figure}[htb]
\centering
\includegraphics[width=12cm]{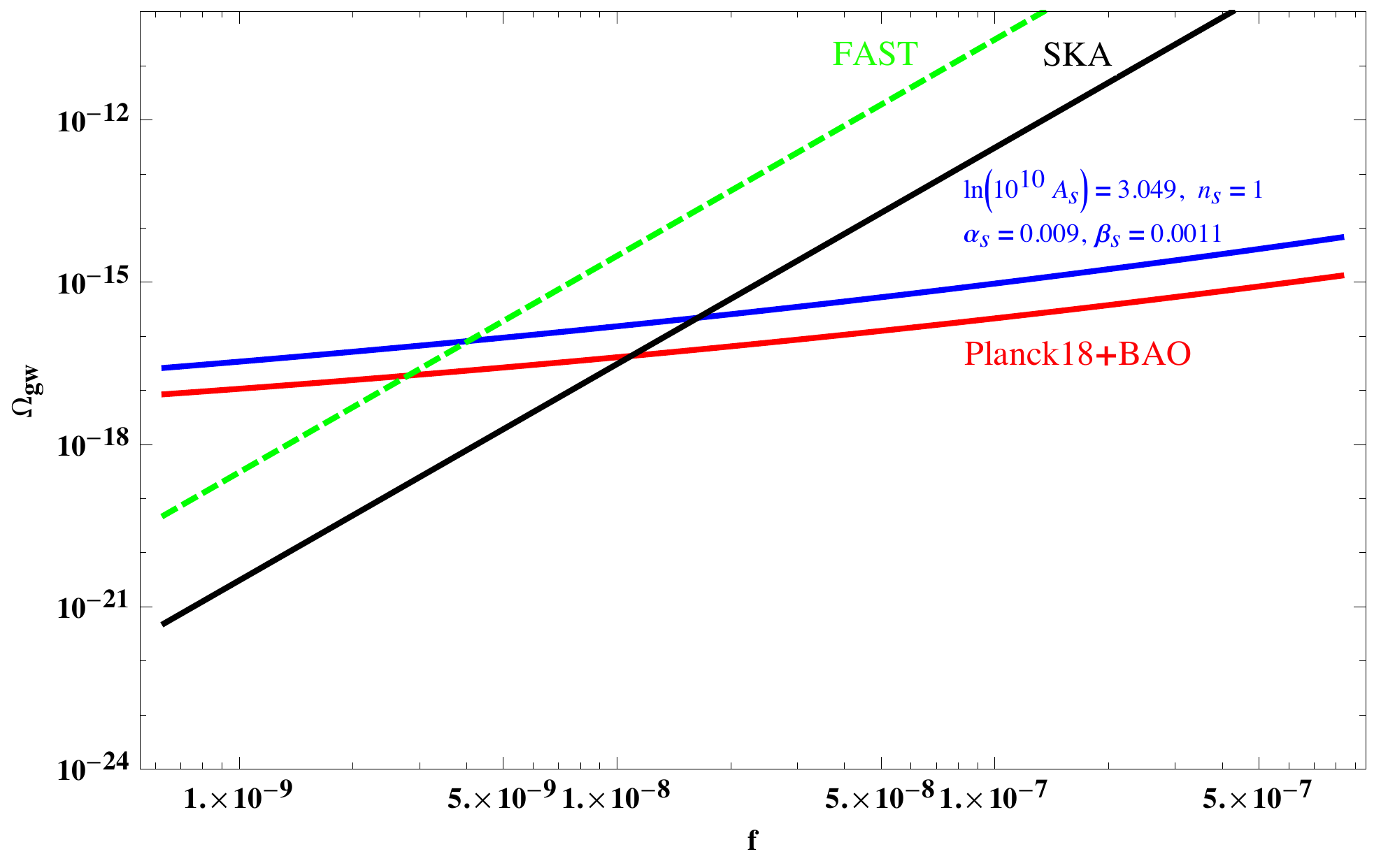}
\caption{The energy density fraction $\Omega_{\mathrm{GW}}$ of the scalar induced gravitational waves in Eq.~(\ref{igw2}) and the sensitivity curves of FAST and SKA detectors. }
\label{figure3}
\end{figure}

Furthermore, we can characterize the scalar fluctuation spectrum in terms of the spectral index $n_s$ and its first two derivatives with respect to $\ln k$
\m
\mathcal{P}_{\zeta}(k)&=&A_s\(\frac{k}{k_*}\)^{n_s-1+\frac{1}{2}\alpha_s\ln(k/k_*)+\frac{1}{6}\beta_s(\ln(k/k_*))^2},
\n
where $\beta_s\equiv{\mathrm{d}^2n_s}/{\mathrm{d}\ln k^2}$ is the running of the running of the spectral index. The power spectrum of scalar induced gravitational waves takes the form
\e
\mathcal{P}_h(\eta, k)=\frac{24Q(n_s, \alpha_s, \beta_s, k)}{(k\eta)^2}A^2_s\(\frac{k}{k_*}\)^{2\[{n_s-1+\frac{1}{2}\alpha_s\ln(k/k_*)+\frac{1}{6}\beta_s(\ln(k/k_*))^2}\]},
\q
where the overall coefficient is
\m
Q(n_s, \alpha_s, \beta_s, k)=&&  \frac{1}{12}\int_0^\infty\mathrm{d}v\int_{\vert1-v\vert}^{1+v}\mathrm{d}u \Bigg(\frac{4v^2-(1+v^2-u^2)^2}{4vu}\Bigg)^2 \Bigg(\frac{3(u^2+v^2-3)}{4u^3v^3x}\Bigg)^2\\ \nonumber
&&\Bigg(\Big(-4uv+(u^2+v^2-3)\log\left|\frac{3-(u+v)^2}{3-(u-v)^2}\right|\Big)^2+\pi^2(u^2+v^2-3)^2\Theta(u+v-\sqrt{3})\Bigg)\\ \nonumber
&&\Bigg(\frac{k}{k_*}\Bigg)^{\frac{1}{2}\alpha_s\ln(uv)+\frac{1}{6}\beta_s\Big((\ln{v})^2+2\ln{v}\ln(k/k_*)+(\ln{u})^2+2\ln{u}\ln(k/k_*)\Big)}\Bigg(uv\Bigg)^{n_s-1+\frac{1}{2}\alpha_s\ln(k/k_*)+\frac{1}{6}\beta_s(\ln(k/k_*))^2}\\ \nonumber
&&v^{\frac{1}{2}\alpha_s\ln{v}+\frac{1}{6}\beta_s\Big((\ln{v})^2+2\ln{v}\ln(k/k_*)\Big)}u^{\frac{1}{2}\alpha_s\ln{u}+\frac{1}{6}\beta_s\Big((\ln{u})^2+2\ln{u}\ln(k/k_*)\Big)},
\n
which depends on $n_s$, $\alpha_s$, $\beta_s$ and $k$. The fractional energy density becomes
\e
\Omega_{\mathrm{GW}}(\eta, k)=Q(n_s, \alpha_s, \beta_s, k)A^2_s\(\frac{k}{k_*}\)^{2\[{n_s-1+\frac{1}{2}\alpha_s\ln(k/k_*)+\frac{1}{6}\beta_s(\ln(k/k_*))^2}\]}.  \label{igw2}
\q
According to Planck18+BAO observations: $n_s=0.9647$, $\alpha_s=0.009$ and $\beta_s=0.0011$, the values of the overall coefficient $Q(n_s, \alpha_s, \beta_s, k)$ are shown in Table.~\ref{table2}. The strength of the scalar induced gravitational waves around the frequency of $10^{-10}$ Hz would be of the order $10^{-17}$ which is presented in Fig.~\ref{figure3}. The sensitivity curve of FAST detector and the energy density fraction $\Omega_{\mathrm{GW}}$ of the scalar induced gravitational waves in Eq.~(\ref{igw2}) would intersect around the frequency of $3*10^{-9}$ Hz. So we could expect that the sensitivity curve of FAST leads to distinct constraints on the running of the running of the spectral index, especially the larger parts.

\begin{table*}[htb]
\newcommand{\tabincell}[2]{\begin{tabular}{@{}#1@{}}#2\end{tabular}}
  \centering
  \begin{tabular}{  c |c| c| c}
  \hline
  \hline
    mode & \tabincell{c} {$k/k_*=4.1*10^6$} & \tabincell{c}{$k/k_*=8.2*10^6$} & \tabincell{c}{$k/k_*=10$}\\
  \hline
  $Q(n_s, \alpha_s, \beta_s, k)$ & $224.33$ &$1529.88$  &$1497.93$\\
  \hline
  \end{tabular}
  \caption{The overall coefficient $Q(n_s, \alpha_s, \beta_s, k)$}
  \label{table2}
\end{table*}

\section{Constraints on primordial curvature perturbations from the scalar induced gravitational waves}
\begin{figure}[htb]
\centering
\includegraphics[width=13cm]{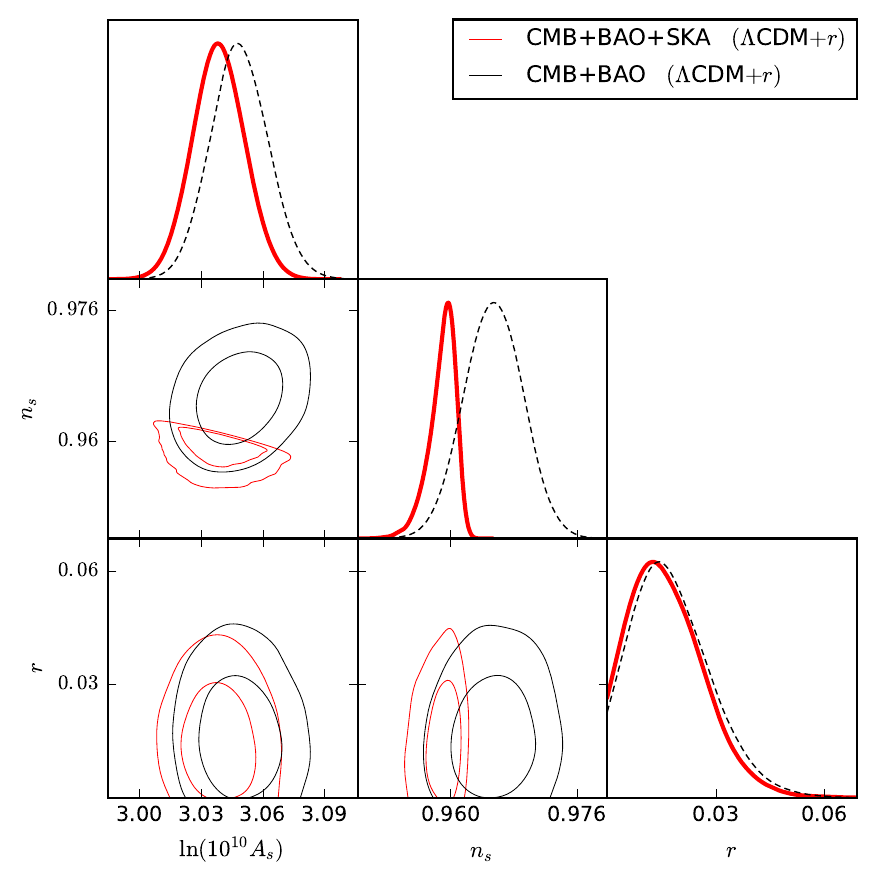}
\caption{The contour plots and the likelihood distributions for the cosmological parameters in the $\Lambda$CDM+$r$ model at the $68\%$ and $95\%$ CL from the combinations of CMB+BAO and CMB+BAO+SKA, respectively. The filled lines in the likelihood distributions are from CMB+BAO+SKA data. The dashed lines in the likelihood distributions are from CMB+BAO data.}
\label{figure4}
\end{figure}

We use the publicly available codes Cosmomc \cite{Lewis:2002ah} to constrain the scalar induced gravitational waves and the power spectrum of primordial curvature perturbations. In the standard $\Lambda$CDM model, the six parameters are the baryon density parameter $\Omega_b h^2$, the cold dark matter density $\Omega_c h^2$, the angular size of the horizon at the last scattering surface $\theta_\text{MC}$, the optical depth $\tau$, the scalar amplitude $A_s$ and the scalar spectral index $n_s$. Usually we introduce a new parameter, namely the tensor-to-scalar ratio $r$, to quantify the tensor amplitude $A_t$ compared to the scalar amplitude $A_s$
at the pivot scale:
\e
r\equiv\frac{A_t}{A_s}.
\q
We extend the standard $\Lambda$CDM model by adding the tensor-to-scalar ratio $r$ and constrain these seven parameters from the combinations of CMB\renewcommand{\thefootnote}{\Roman{footnote}}\footnote{CMB=Planck18+BK18}+BAO, CMB+BAO+FAST, and CMB+BAO+SKA, respectively. Our numerical results are given in Table.~\ref{table3} and Fig.~\ref{figure4}.

\begin{table*}[htb]
\newcommand{\tabincell}[2]{\begin{tabular}{@{}#1@{}}#2\end{tabular}}
  \centering
  \begin{tabular}{  c |c| c| c}
  \hline
  \hline
  Parameter & \tabincell{c} {CMB+BAO} & \tabincell{c}{CMB+BAO+FAST} & \tabincell{c}{CMB+BAO+SKA}\\
  \hline
  $\Omega_bh^2$ & $0.02241\pm0.00013$ &$0.02241\pm0.00013$  &$0.02233\pm0.00013$\\
  $\Omega_ch^2$ &$0.11954\pm{0.00091}$  &$0.11953\pm{0.00095}$  &$0.12056\pm{0.00079}$\\
  $100\theta_{\mathrm{MC}}$ & $1.04099\pm0.00029$   &$1.04100\pm0.00029$   &$1.04088^{+0.00027}_{-0.00028}$\\
  $\tau$ &  $0.0567^{+0.0070}_{-0.0071}$ &$0.0566^{+0.0069}_{-0.0076}$  &$0.0505^{+0.0063}_{-0.0062}$\\
  $\ln\(10^{10}A_s\)$  & $3.049\pm0.014$   &$3.048^{+0.014}_{-0.015}$   &$3.038\pm0.013$\\
  $n_s$ & $0.9654\pm{0.0037}$ &$0.9654\pm0.0038$   &$0.9589^{+0.0021}_{-0.0011}$\\
  $r_{0.05}$  ($95\%$ CL) &$<0.038$ &$<0.037$  &$<0.035$\\
  \hline
  \end{tabular}
  \caption{The $68\%$ limits on the cosmological parameters in the $\Lambda$CDM+$r$ model from the combinations of CMB+BAO, CMB+BAO+FAST and CMB+BAO+SKA, respectively.}
  \label{table3}
\end{table*}

We see that the constraints on the power spectrum of primordial curvature perturbations are affected by adding the upper limit of scalar induced gravitational waves from SKA project. If there is no detection of scalar induced gravitational waves from SKA project, the scalar amplitude $A_s$ and  the spectral index $n_s$ of the power-law spectrum are smaller than the constraints from CMB+BAO data which are obvious in Fig.~\ref{figure4}. The larger parts are cut off as we expect. We also consider the effects of LIGO, Virgo, LISA, IPTA and FAST detectors. While these detectors do not modify the constraints on the scalar amplitude and the spectral index from CMB+BAO data, the sensitivity curve of CMB+BAO are totally within the upper limits of scalar induced gravitational waves.

\begin{figure}[htb]
\centering
\includegraphics[width=13cm]{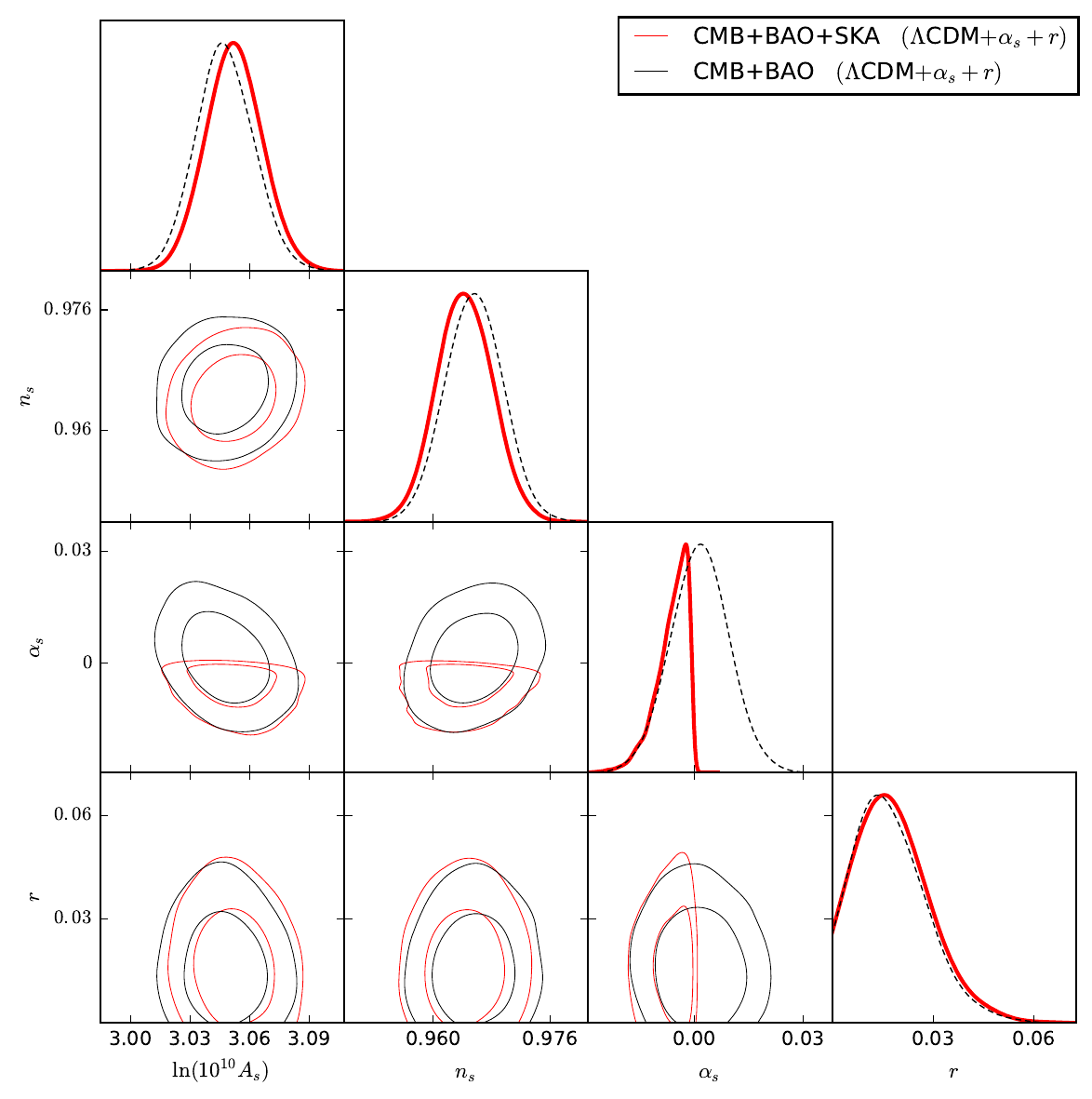}
\caption{The contour plots and the likelihood distributions for the cosmological parameters in the $\Lambda$CDM+$\alpha_s$+$r$ model at the $68\%$ and $95\%$ CL from the combinations of CMB+BAO and CMB+BAO+SKA, respectively. The filled lines in the likelihood distributions are from CMB+BAO+SKA data. The dashed lines in the likelihood distributions are from CMB+BAO data.}
\label{figure5}
\end{figure}

\begin{figure}[htb]
\centering
\includegraphics[width=12.5cm]{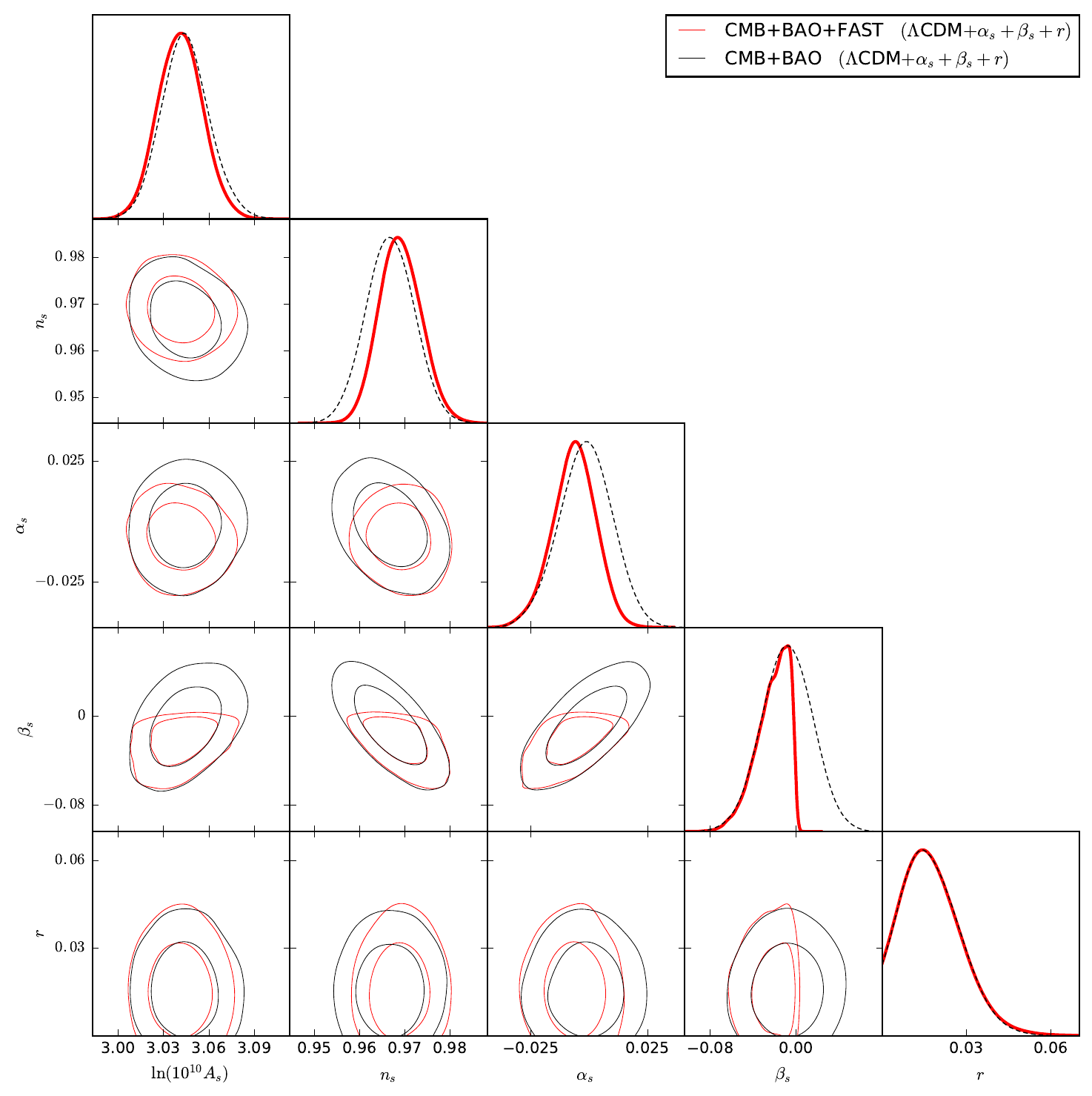}
\caption{The contour plots and the likelihood distributions for the cosmological parameters in the $\Lambda$CDM+$\alpha_s$+$\beta_s$+$r$ model at the $68\%$ and $95\%$ CL from the combinations of CMB+BAO and CMB+BAO+FAST, respectively. The filled lines in the likelihood distributions are from CMB+BAO+FAST data. The dashed lines in the likelihood distributions are from CMB+BAO data.}
\label{figure6}
\end{figure}

\begin{figure}[htb]
\centering
\includegraphics[width=12.5cm]{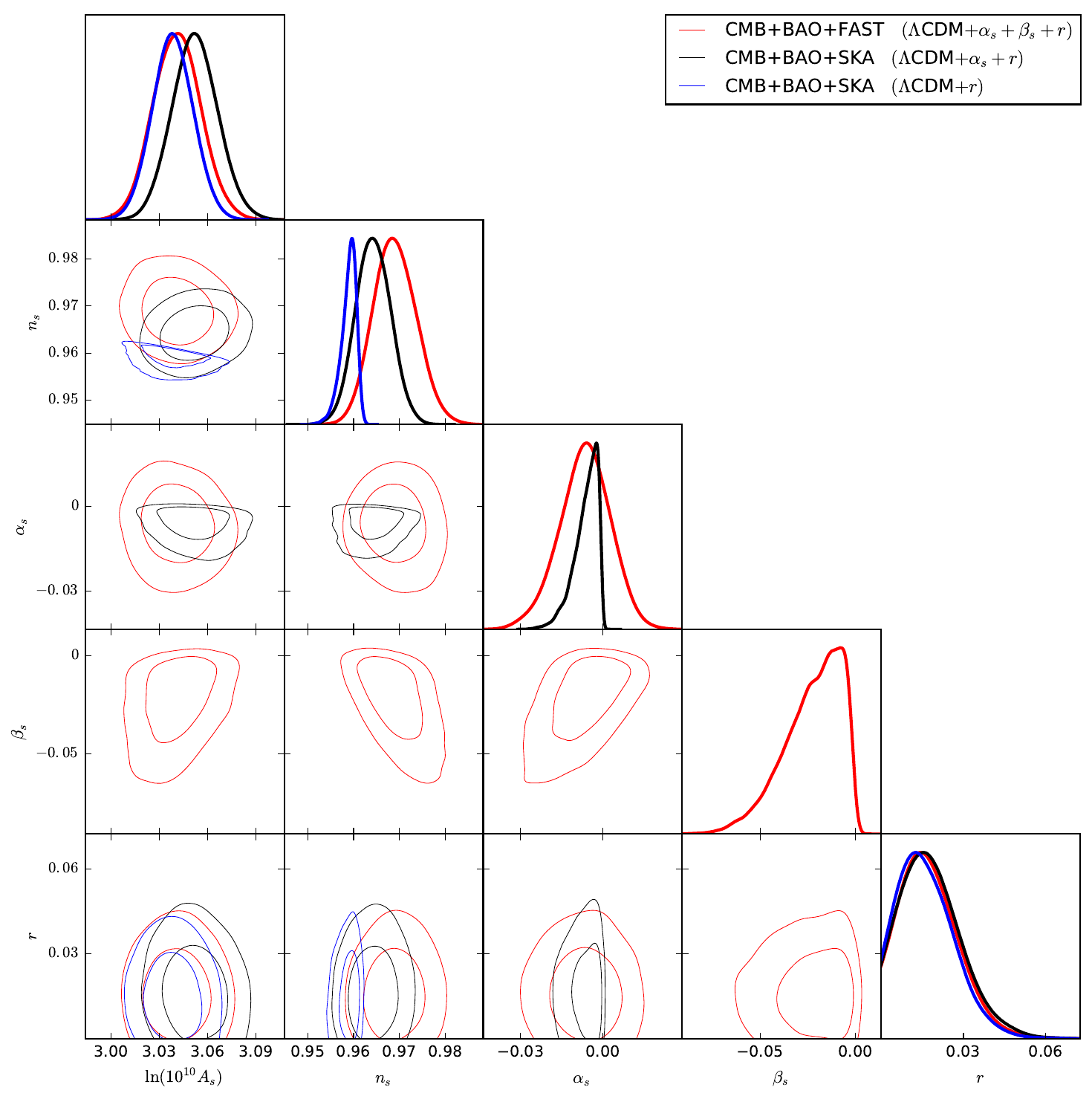}
\caption{The contour plots and the likelihood distributions for the cosmological parameters at the $68\%$ and $95\%$ CL in the $\Lambda$CDM+$r$ model and the $\Lambda$CDM+$\alpha_s$+$r$ model  from the combinations of CMB+BAO+SKA, and in the $\Lambda$CDM+$\alpha_s$+$\beta_s$+$r$ model from the combinations of CMB+BAO+FAST.}
\label{figure7}
\end{figure}

Adding the running of the spectral index $\alpha_s$ and the running of the running of the spectral index $\beta_s$, we can obtain the constraints of the $\Lambda$CDM+$\alpha_s$+$r$ model and the $\Lambda$CDM+$\alpha_s$+$\beta_s$+$r$ model from the combinations of CMB+BAO, CMB+BAO+FAST, and CMB+BAO+SKA, respectively. Our numerical results are given in Table.~\ref{table4}, Table.~\ref{table5} and Fig.~\ref{figure5} to Fig.~\ref{figure7}.

\begin{table*}[htb]
\newcommand{\tabincell}[2]{\begin{tabular}{@{}#1@{}}#2\end{tabular}}
  \centering
  \begin{tabular}{  c |c| c}
  \hline
  \hline
  Parameter & \tabincell{c} {CMB+BAO}  & \tabincell{c}{CMB+BAO+SKA}\\
  \hline
  $\Omega_bh^2$ & $0.02239\pm0.00015$   &$0.02245\pm0.00014$\\
  $\Omega_ch^2$ &$0.11955_{-0.00094}^{+0.00093}$    &$0.11960_{-0.00094}^{+0.00092}$\\
  $100\theta_{\mathrm{MC}}$ & $1.04099_{-0.00029}^{+0.00030}$      &$1.04100_{-0.00028}^{+0.00029}$\\
  $\tau$ &  $0.0563^{+0.0069}_{-0.0078}$  &$0.0578^{+0.0070}_{-0.0076}$\\
  $\ln\(10^{10}A_s\)$  & $3.048_{-0.014}^{+0.015}$     &$3.052\pm0.014$\\
  $n_s$ & $0.9656\pm{0.0039}$   &$0.9643\pm{0.0038}$\\
  $\alpha_s$ &$0.0015_{-0.0080}^{+0.0079}$ &$-0.0062_{-0.0023}^{+0.0056}$\\
  $r_{0.05}$  ($95\%$ CL) &$<0.037$   &$<0.039$\\
  \hline
  \end{tabular}
  \caption{The $68\%$ limits on the cosmological parameters in the $\Lambda$CDM+$\alpha_s$+$r$ model from the combinations of CMB+BAO and CMB+BAO+SKA, respectively.}
  \label{table4}
\end{table*}

\begin{table*}[htb]
\newcommand{\tabincell}[2]{\begin{tabular}{@{}#1@{}}#2\end{tabular}}
  \centering
  \begin{tabular}{  c |c| c}
  \hline
  \hline
  Parameter & \tabincell{c} {CMB+BAO} & \tabincell{c}{CMB+BAO+FAST}\\
  \hline
  $\Omega_bh^2$ & $0.02241\pm0.00015$ &$0.02243\pm0.00014$\\
  $\Omega_ch^2$ &$0.11950\pm{0.00096}$  &$0.11936_{-0.00094}^{+0.00093}$\\
  $100\theta_{\mathrm{MC}}$ & $1.04098\pm0.00029$    &$1.04099\pm0.00029$\\
  $\tau$ &  $0.0546^{+0.0075}_{-0.0088}$  &$0.0528^{+0.0073}_{-0.0072}$\\
  $\ln\(10^{10}A_s\)$  & $3.044^{+0.015}_{-0.017}$      &$3.041\pm{0.015}$\\
  $n_s$ & $0.9668\pm{0.0053}$    &$0.9690^{+0.0046}_{-0.0050}$\\
  $\alpha_s$ &$-0.0016\pm0.011$ &$-0.0065^{+0.010}_{-0.009}$\\
  $\beta_s$ &$-0.0083\pm0.023$ &$-0.0223^{+0.020}_{-0.009}$\\
  $r_{0.05}$  ($95\%$ CL) &$<0.036$   &$<0.037$\\
  \hline
  \end{tabular}
  \caption{The $68\%$ limits on the cosmological parameters in the $\Lambda$CDM+$\alpha_s$+$\beta_s$+$r$ model from the combinations of CMB+BAO and CMB+BAO+FAST, respectively.}
  \label{table5}
\end{table*}

In the $\Lambda$CDM+$\alpha_s$+$r$ model, we find that the mean value of the scalar amplitude $A_s$ shifts to an upper value, while the mean values of the spectral index $n_s$ and the running of the spectral index $\alpha_s$ shift to lower values when SKA is added to the CMB+BAO data. In the $\Lambda$CDM+$\alpha_s$+$\beta_s$+$r$ model, the mean value of the scalar amplitude $A_s$ shifts to a lower value. The mean value of the spectral index $n_s$ shifts to an upper value and the mean values of $\{\alpha_s, \beta_s\}$ turn to lower values when FAST is added to the CMB+BAO data. The upper limit of scalar induced gravitational waves from SKA or FAST project changes the contours and likelihoods from the black ones to the red ones in Fig.~\ref{figure4} to Fig.~\ref{figure6}. Here, we can compare the constraints of $\Lambda$CDM+$r$ model, $\Lambda$CDM+$\alpha_s$+$r$ model and $\Lambda$CDM+$\alpha_s$+$\beta_s$+$r$ model from the combinations of CMB+BAO+SKA and CMB+BAO+FAST in Fig.~\ref{figure7}. When we consider the running of the running of the spectral index $\beta_s$, the index factor becomes more sensitive and the scalar amplitude becomes less sensitive to the upper limit of scalar induced gravitational waves. Except the scalar induced gravitational waves, we can evaluate upper bounds on the scalar spectrum from primordial black holes \cite{Alabidi:2012ex,Li:2018iwg,Dalianis:2018ymb}.

Besides, the gravitational waves detectors are able to actually constrain primordial gravitational waves. The power spectra of the tensor perturbations are parameterized as
\m
\mathcal{P}_t(k)&=&A_t\(\frac{k}{k_*}\)^{n_t},
\n
where $n_t=-r/8$ is the consistency relation in the single-field slow-roll inflation model. According to CMB+BAO observations, we would expect the primordial gravitational wave spectrum to dominate over the scalar induced gravitational waves. The primordial gravitational waves could be constrained by the gravitational waves observations \cite{Campeti:2020xwn,Li:2021uvn, Li:2021scb, Li:2019vlb}.

\section{summary}
In this paper, we constrain the fractional energy density of scalar induced gravitational waves from gravitational waves observations. If there is no detection of the scalar induced gravitational waves, the fractional energy density of scalar induced gravitational waves is confined by some upper limits. Depends on these upper limits, we can obtain the constraints on the power spectrum of the primordial curvature perturbations. For a power-law scalar power spectrum, the constraints on the power spectrum are affected by adding the upper limit of scalar induced gravitational waves from SKA project. In the $\Lambda$CDM+$r$ model, the mean values of the scalar amplitude and the spectral index shift to lower values when SKA is added to the CMB+BAO data. We also consider the effects of LIGO, Virgo, LISA, IPTA and FAST detectors, while the constraints from CMB+BAO are totally within their upper limits of scalar induced gravitational waves. Furthermore, we characterize the scalar fluctuation spectrum in terms of the spectral index $n_s$ and its first two derivatives. We calculate corresponding power spectrum of scalar induced gravitational waves theoretically and give the constraints on the running of the spectral index and the running of the running of the spectral index.

\noindent {\bf Acknowledgments}.
This work is supported by Natural Science Foundation of Shandong Province (grant No. ZR2021QA073) and Research Start-up Fund of QUST (grant No. 1203043003587).



\end{document}